# Chapter 3

# The IITM Earth System Model (IITM ESM)

**Lead Author:** *R. Krishnan and P. Swapna*
**Co-authors:** *Ayantika Dey Choudhury, Sandeep Narayansetti, A.G. Prajeesh, Manmeet Singh, Aditi Modi, Roxy Mathew, Ramesh Vellore, J. Jyoti, T.P. Sabin, J. Sanjay, Sandip Ingle*

1. Introduction
2. A brief description of the IITM-ESM
3. Salient features of the IITM-ESMv2
4. References

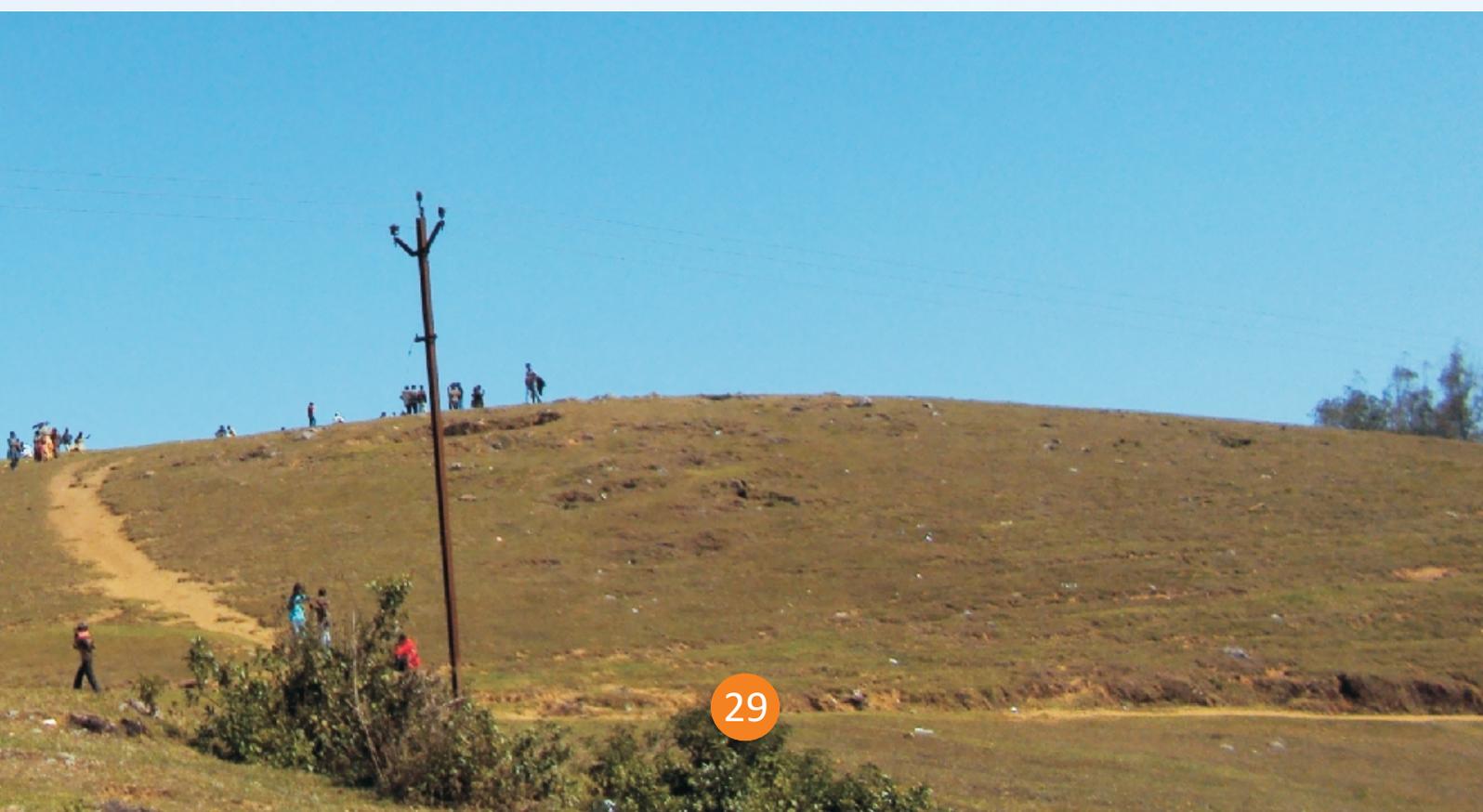

## 1. Introduction

Earth System Models (ESM) are important tools that allow us to understand and quantify the physical, chemical & biological mechanisms governing the rates of change of elements of the Earth System, comprising of the atmosphere, ocean, land, cryosphere and biosphere (terrestrial and marine) and related components. ESMs are essentially coupled numerical models which incorporate processes within and across the different Earth system components and are expressed as set of mathematical equations. ESMs are useful for enhancing our fundamental understanding of the climate system, its multi-scale variability, global and regional climatic phenomena and making projections of future climate change. In this chapter, we briefly describe the salient aspects of the Indian Institute of Tropical Meteorology ESM (IITM ESM), that has been developed recently at the IITM, Pune, India, for investigating long-term climate variability and change with focus on the South Asian monsoon.

Observations of the climate system based on direct measurements and remote sensing from satellites and other platforms indicate that the warming of the climate system has been unequivocal since the 1950s, many of the observed changes are unprecedented over decades to millennia (Stocker et al. 2013). Long-term climate model simulations, that took part in the Intergovernmental Panel for Climate Change (IPCC) fifth assessment report (AR5), provide very high confidence in interpreting the observed global-mean surface temperature trends during the post-1950s; as well as the human influence on the climate system which is clearly evident from the increasing concentration of greenhouse gases (GHG) in the atmosphere, positive radiative forcing and the observed global warming (Stocker et al. 2013). On the other hand, there are major challenges in comprehending the impacts of climate change at regional levels. For example, the 20th century simulations and future projections of the South Asian monsoon rainfall based on the IPCC models exhibit a wide range of variations and uncertainties *[Ref: Turner and Annamalai, 2012; Sharmila et al., 2014; Saha et al 2015; Krishnan et al. 2016],* which pose huge challenges to policy makers and development of adaptation strategies.

The Coupled Modeling Intercomparison Project (CMIP3; [*Meehl et al.*, 2007], CMIP5; [*Taylor et al.*, 2012]) coordinated by the World Climate Research Programme (WCRP) form the basis of the climate projections in the Intergovernmental Panel on Climate Change (IPCC) Assessment Reports. The CMIP models have convincingly demonstrated the role of anthropogenic forcing on the rising trend of global mean temperatures [*Flato et al.*, 2013], however there are challenges in quantifying the response of regional monsoon precipitation to climate change [*eg., Turner and Annamalai*, 2012; *Sperber et al.*, 2013; *Kitoh*, 2017].

The underlying philosophy behind the IITM ESM is based on developing a global modeling framework to address the science of climate change, including detection, attribution and future projections of global

### Highlights

- **IITM-ESMv1:** *Successful development of the first version of IITM ESM at CCCR, IITM, Pune by transforming a seasonal prediction model (CFSv2) into a long term climate model (Ref: Swapna et al. 2015). This development was achieved by incorporating a new ocean model component (MOM4p1, including ocean biogeochemistry) in CFSv2. Major improvements in the IITM-ESM relative to CFSv2 include:*

    - *Significant reduction of cold bias of global mean sea surface temperature (SST) by ~0.8°C*
    - *Robust simulations of drivers of natural modes of global climate variability [eg., El Nino/Southern Oscillation (ENSO) and Pacific Decadal Oscillation (PDO)] well-captured in IITM-ESMv1*
    - *Teleconnections between ENSO and the Indian monsoon rainfall well captured in IITM-ESMv1*

- **IITM-ESMv2:** *Successful development of the second version of IITM ESM at CCCR, IITM, Pune by incorporating various refinements and improvements in the first version. The IITM-ESMv2 is a radiatively balanced global climate modeling framework that has the capability to address key scientific questions relating to long-term climate change.*

- **IITM-ESMv2 to participate in CMIP6:** *The IITM-ESMv2 would be first climate model from India to contribute to the Coupled Model Intercomparison Project Sixth Phase (CMIP6) for the IPCC sixth assessment report (Ar6).*

climate, with special emphasis on the South Asian monsoon. With this view, the first version (IITM ESM version 1) was developed by transforming a state-of-the-art seasonal prediction model, Climate Forecast System version 2 (CFSv2, [*Saha et al.*, 2010], into a model suitable for long-term climate [*Swapna et al. 2015*]. Subsequently an updated version of the IITM Earth System Model (IITM ESM version 2) has been developed at the CCCR-IITM, Pune, by incorporating various refinements leading to a radiatively balanced global climate modeling framework appropriate for addressing the science of climate change.

## 2. A brief description of the IITM-ESM

The IITM-ESMv2 configuration includes *(a)* A atmosphere general circulation model [global spectral model with triangular truncation of 62 waves (T62, grid size ~200 km) and 64 vertical levels with top model layer extending up to 0.2 hPa *(b)* A global ocean component based on the Modular Ocean Model Version 4p1 (MOM4p1) (*Griffies*, 2009) having a zonal resolution of ~100 km and the meridional resolution ~ 35 km between 10°S and 10°N and coarser grid ~100 km poleward of 30° latitude in both Hemispheres and with 50 levels in the vertical *(c)* A land surface model (Noah LSM) with four layers *(d)* A dynamical sea-ice model known as the Sea Ice Simulator (SIS) (*Winton*, 2000). A schematic of the IITM-ESMv2 is shown in Figure 3.1. Here we present highlights of the IITM-ESMv2 simulations, which include multi-century runs corresponding to the pre-industrial and future climatic conditions following the Coupled Model Intercomparison Project Phase 6 (CMIP6) protocols. The IITM-ESMv2 would be the first climate model from India contributing to the CMIP6 experiments for the Intergovernmental Panel for Climate Change (IPCC) sixth assessment report (AR6) to be released in 2021.

## 3. Salient features of the IITM-ESMv2

The IITM-ESMv2 has capabilities for conducting long-term climate change studies. The salient features of the IITM-ESMv2 include:

a. ***Radiatively balanced framework:*** *Global balance of net radiative fluxes at the top-of-the-atmosphere (TOA), surface of the Earth (surface)*

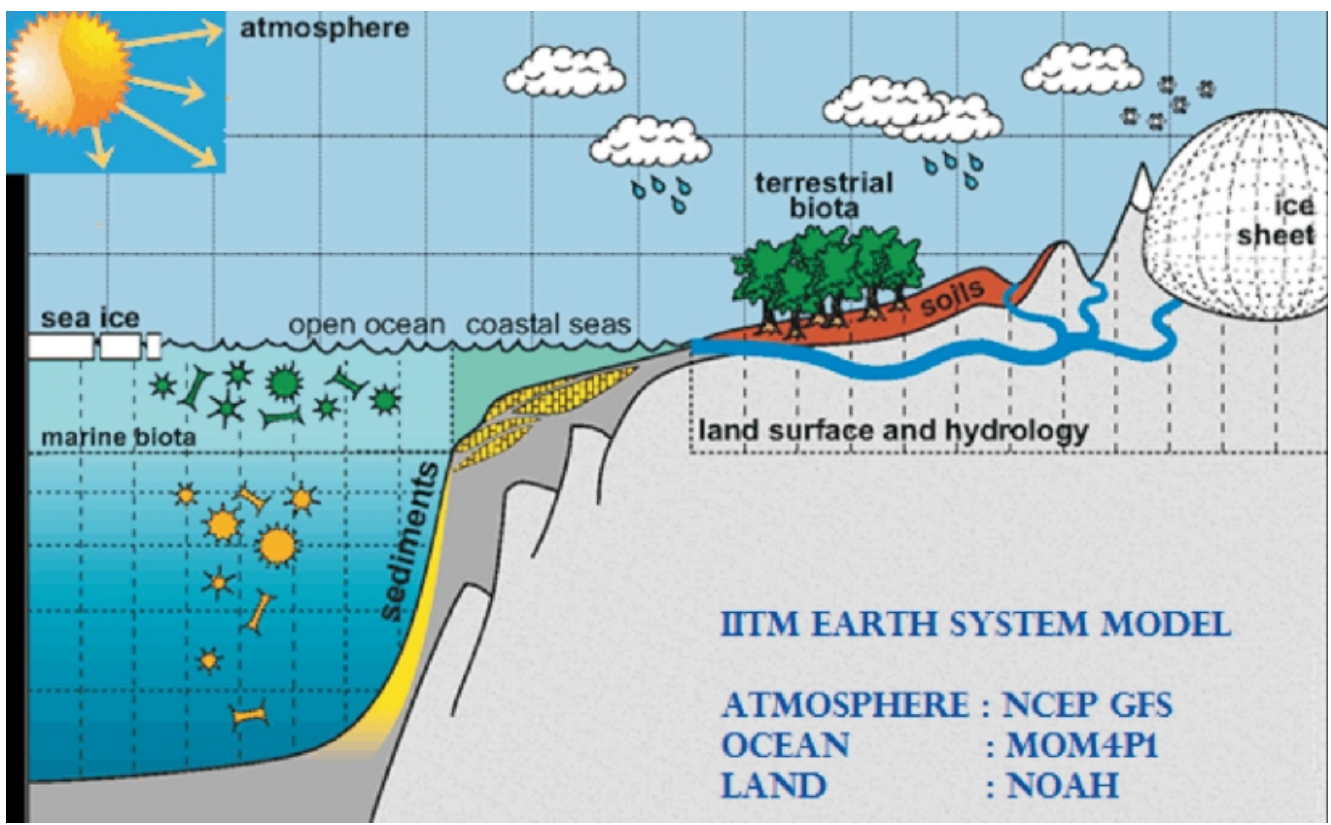

**Figure 3.1** Schematic showing different components of the IITM Earth System Model.

and atmosphere. This is an essential criterion to be met by climate models in order to make reliable assessments of the impacts of the radiative effects from anthropogenic forcing (eg., GHG, aerosols, …) on the climate system (Figure.3.2). The IITM-ESMv2 exhibits fidelity in capturing the global and tropical climatic features (Figures.3.3-3.4)

b. **South Asian Monsoon:** Improved simulation of time-mean monsoon precipitation over South Asia (Figure.3.5). Given that the Indian Monsoon (a.k.a South Asian Monsoon) is the lifeline of the regional socio-economic activities, there is strong emphasis at CCCR, IITM to make reliable assessments of changes in the Indian monsoon precipitation under climate change. Keeping this in view, the model developmental efforts at CCCR, IITM focused on refining the representations of both the global climate as well as the regional monsoon phenomenon in the IITM-ESMv2.

c. **Marine biogeochemistry:** IITM-ESMv2 incorporates interactive ocean biogeochemistry and ecosystem processes. This allows us to investigate the impacts of climate variability and climate change on marine primary productivity and mechanisms that control the ocean carbon cycle (Figure.3.6).

d. **Polar sea-ice distribution:** Sea-ice distribution in polar region is an important component of the climate system. The IITM-ESMv2 shows a realistic representation of time-mean distribution of polar sea-ice, as compared to IITM-ESMv1 which severely underestimated the sea-ice cover (Figure 3.7).

e. **Atlantic Meridional Overturning Circulation (AMOC):** A significant outcome of improving polar sea-ice distribution is manifested by a realistic simulation of AMOC in the IITM-ESM. The AMOC, which is a deep ocean circulation driven by large-scale density (temperature & salinity) gradients in the ocean interior, is a major driver of global climate variability. Global warming impacts are projected to affect the strength of AMOC through their influence on polar ice caps (both over sea and land). These impacts can affect the global climate as well as the Asian monsoon by altering the atmospheric and oceanic circulation patterns.

f. **Aerosol forcing:** IITM-ESMv2 incorporates the radiative effects of aerosols, both natural (eg., dust, sea-salt, volcanic emissions…) and anthropogenic (sulfate, nitrate, organic carbon, black carbon, …), on the climate system. Atmospheric aerosols affect climate through scattering and absorption of the incoming solar radiation (direct effect) and through modification of cloud properties (indirect effect). Rapid industrialization during the last 5-6 decades has increased atmospheric aerosol loading [IPCC 2013]. Recent studies have pointed to the role of anthropogenic aerosol forcing on radiation, monsoon rainfall and regional climate [eg., Ramanathan et al. 2005, Bollasina et al. 2011, Krishnan et al. 2016]. Unlike GHGs, the space-time variability of aerosols is large. Furthermore, monsoon precipitation variability over South Asia *is* strongly influenced by monsoon internal dynamics. Therefore, reliable attributions of aerosol forcing on regional precipitation changes have been challenging. The IITM-ESMv2 serves as a valuable tool to address these scientific problems.

g. **Land use and land cover changes:** The IITM-ESMv2 has capabilities to address the effects of land use and land cover changes (LULC) on the climate system. This is of considerable interest to the Asian region which has undergone major changes in forest cover, agricultural land and vegetation types since pre-industrial times. The IITM-ESMv2 provides a great opportunity to investigate the role of LULC on the regional monsoon precipitation pattern.

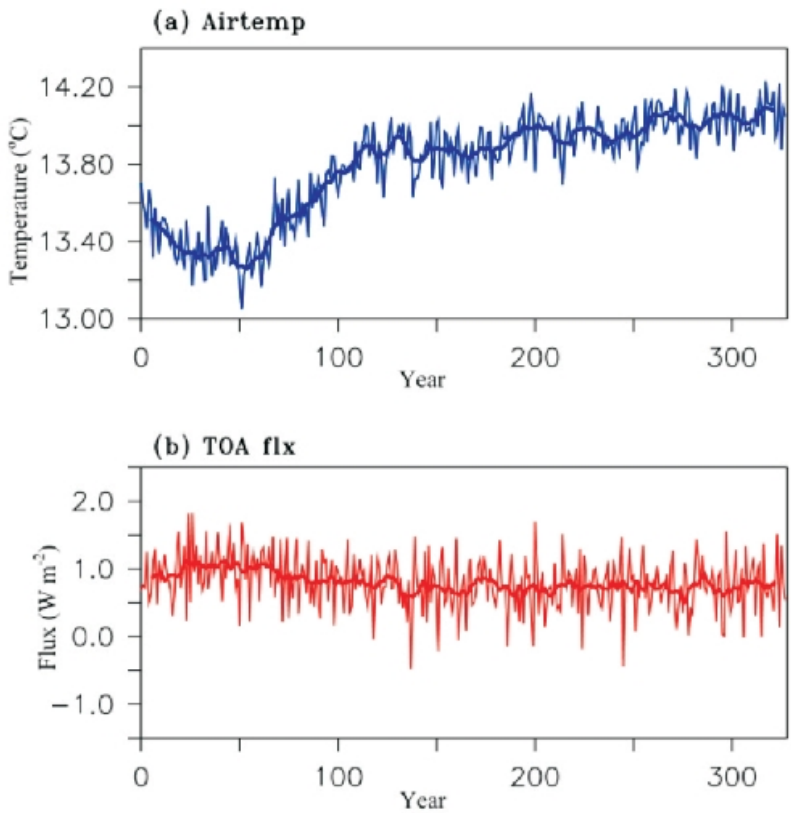

*Figure 3.2* Time-series plots from the Pre-Industrial (PI) control simulation of IITM ESM2 *(a)* Global mean surface air-temperature (°C) *(b)* Net radiation flux at the top-of-atmosphere (TOA). The PI control experiment is a multi-century simulation uses GHG, aerosols, land-use & land-cover and other forcing corresponding to 1850. The PI control experiment of IITM ESM2 has completed more than 300 years of simulation on the IITM HPC and will continue for a few more centuries. It can be noticed that the time-series tend towards quasi-equilibrium with mean values of global mean surface-air-temperature ~14°C and TOA net radiation flux ~ 0.8 Wm$^{-2}$.

*Figure 3.3* Spatial map of the observed climatology *(a)* Annual mean sea surface temperature (SST °C) *(b)* Mean precipitation (mm day$^{-1}$) during the June-July-August-September (JJAS) boreal summer monsoon season. The SST data is based on Hadley Centre dataset (HadISST, Rayer et al. 2003) and the precipitation data is estimated from the Tropical Rainfall Measurement Mission (TRMM, Huffman et al., 2010) satellite. The warm pool region in the tropical eastern Indian Ocean and western Pacific Ocean are associated with SST > 29°C, while cooler SSTs prevail in the eastern equatorial Pacific (< 22°C) giving rise to a strong east-west gradient in the tropical Pacific Ocean. Cold SSTs (< 18°C) are seen in the extra-tropical oceanic areas. The boreal summer monsoon precipitation is dominated by tropical precipitation over the Indian subcontinent, Southeast and East Asia, Tropical Eastern Indian Ocean and Western Pacific, the Inter-Tropical-Convergence-Zone (ITCZ) over the equatorial Pacific and Atlantic, Central and Latin America, equatorial Africa. Extra-tropical precipitation is seen over the East Asian region including eastern China, Korea and Japan, and the east coast of North America.

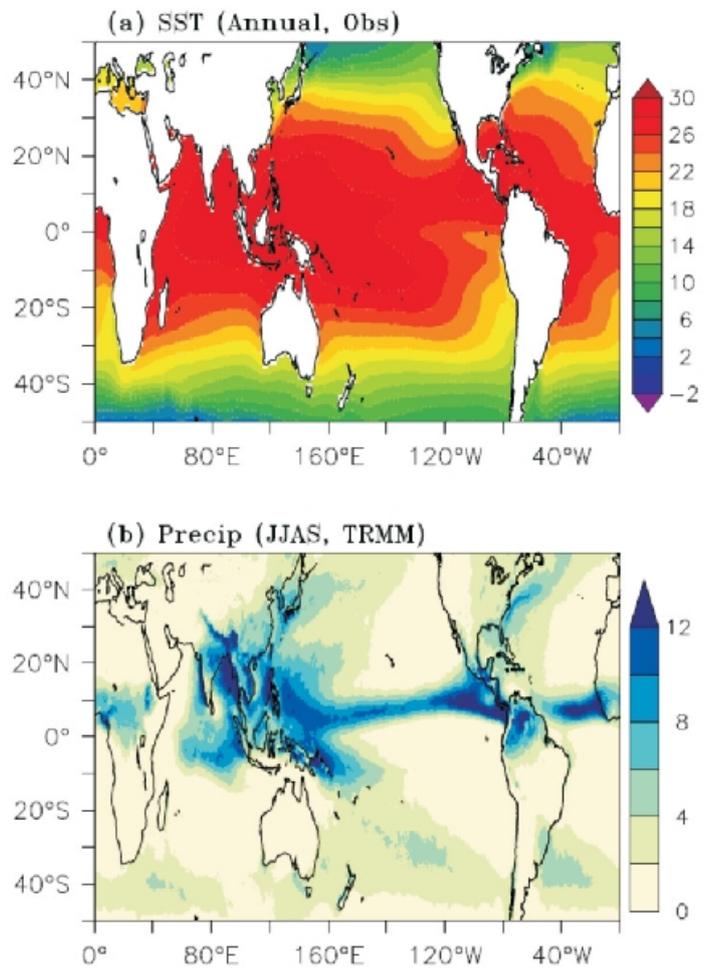

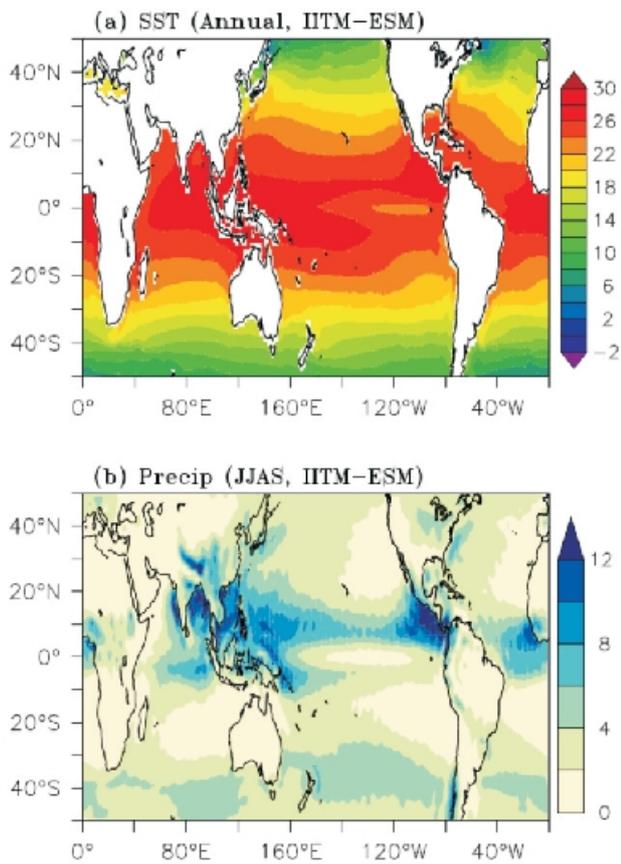

*Figure 3.4* Spatial map of climatological (a) Mean SST (°C) (b) Mean JJAS precipitation (mm day$^{-1}$) simulated by the IITM-ESMv2 from the PI control simulation. The mean values are based on the last 100 years of the PI Control simulation. The broad spatial patterns of the simulated SST and rainfall are consistent with the observed patterns. The magnitude of the simulated SST and rainfall in the tropical Indo-Pacific warm pool are underestimated as compared to observations.

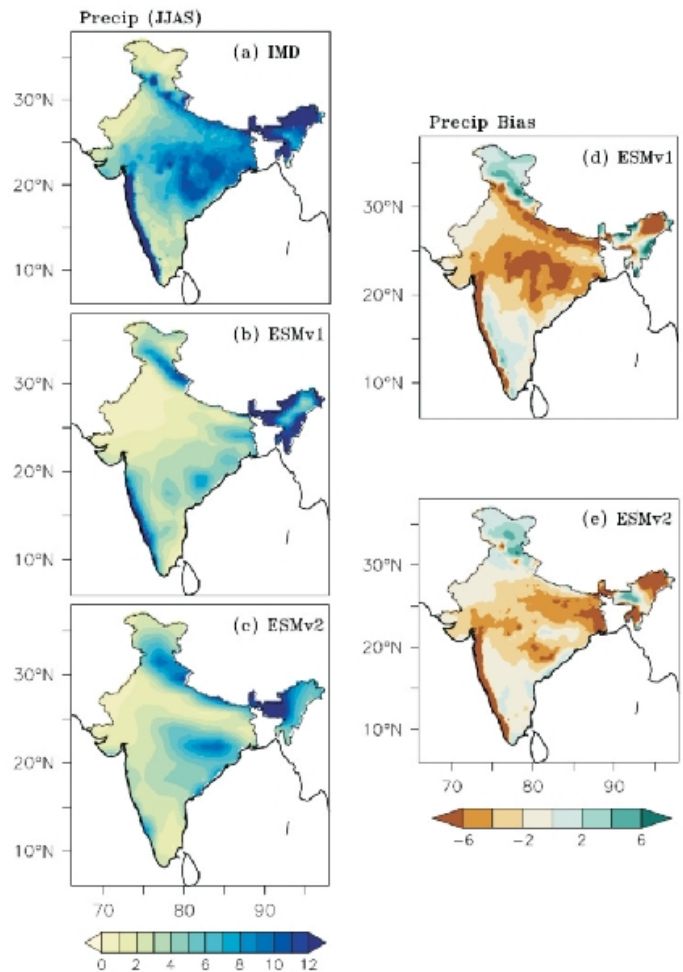

*Figure 3.5* Spatial map of mean boreal summer monsoon (JJAS) precipitation (mm day$^{-1}$) over India **(a)** Observed precipitation from the India Meteorological Department (IMD) based on the Pai et al. (2015) dataset **(b)** Simulated precipitation from IITM-ESMv1 **(c)** Simulated precipitation from IITM-ESMv2 **(d)** Difference (IITM-ESMv1 minus IMD observation) shows the systematic bias in the IITM-ESMv1 simulation **(e)** Difference (IITM-ESMv2 minus IMD observation) shows the systematic bias in the IITM-ESMv2 simulation. It can be noticed that the IITM-ESMv1 has a large dry bias over north-central India. The magnitude of the negative bias in precipitation has been reduced in the IITM-ESMv2 simulation.

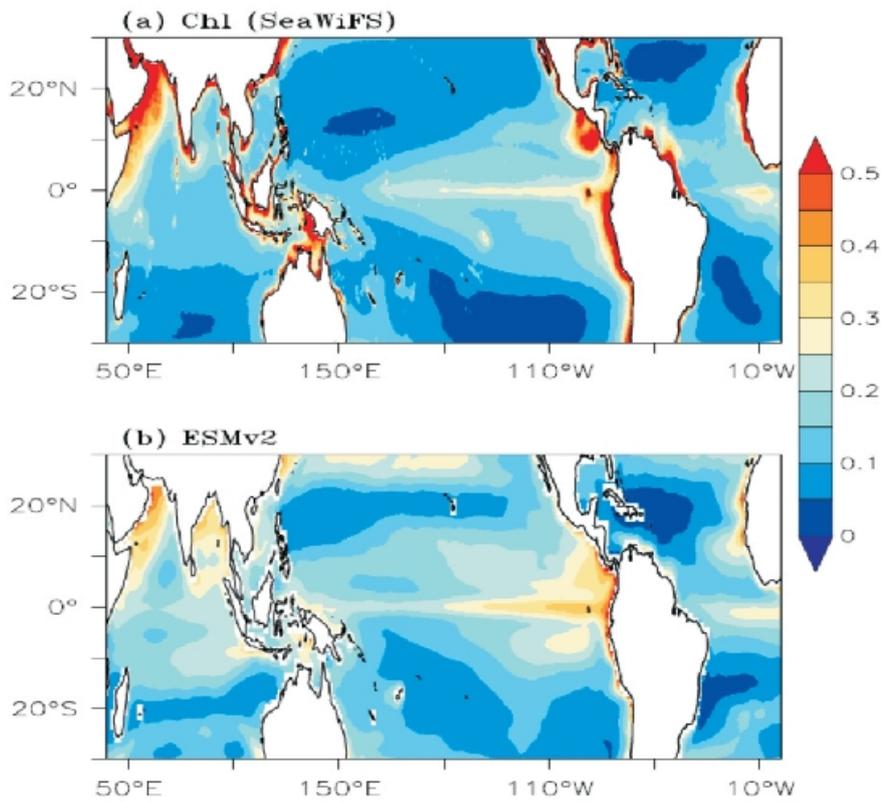

*Figure 3.6* Spatial maps of climatological mean chlorophyll concentration (mg m$^{-3}$) **(a)** Satellite estimates (SeaWifs) **(b)** IITM-ESMv2 simulation. High chlorophyll concentrations off the Somali Coast and Arabian Sea, and the eastern Pacific are associated with oceanic upwelling. The IITM-ESMv2 captures the high-chlorophyll patterns in the northern Indian Ocean and eastern Pacific, although the magnitudes are somewhat underestimated in the Arabian Sea. Also the simulated chlorophyll in the Pacific Ocean extends far westward from the Eastern to the Central Pacific, as compared to observations.

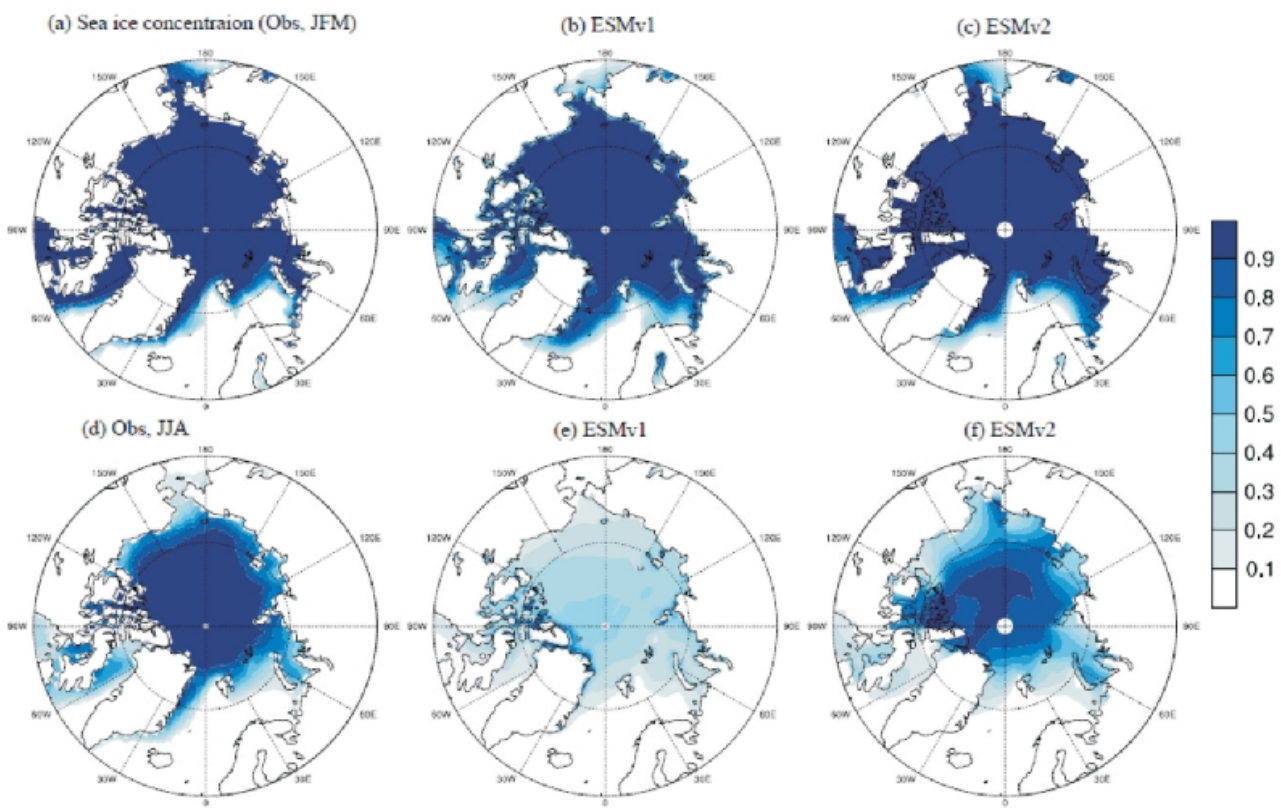

*Figure 3.7* Spatial map of mean sea-ice concentration (%) over the Arctic from observations (Hadley Centre) and IITM-ESMv1 and IITM-ESMv2 simulations **(a-c)** Winter and early Spring (Jan-Feb-Mar) **(d-f)** Northern Summer (Jun-Jul-Aug). The IITM-ESMv1 shows severely depleted sea-ice concentration during the northern summer as compared to observations. This bias in the sea-ice cover simulation is significantly remedied in IITM-ESMv2.

### Box 3.1 Ongoing work and future plans

⊙ Pre-Industrial (PI) control experiment of IITM-ESMv2 following the CMIP6 protocols is ongoing using the High Performance Computing (HPC) at IITM, Pune. This is a multi-century (> 500 years) simulation based on the pre-industrial conditions of GHG, aerosols, land-use & land-cover and other forcing. In addition special experiments are planned to understand long-term climate variability and trends during the 20$^{th}$ century and future projections (Figures 3.8, 3.9) as part of CMIP6 (please see list below). Completion of all the experiments, analysis of results, assessment and scientific publications of the CMIP6 simulations are targeted from 2017 through 2020.

- ⊙ PI Control (> 500 years)
- ⊙ Historical simulation (1850-2014)
- ⊙ AMIP Historical simulation forced with observed SST and sea-ice (1979-2014)
- ⊙ Transient climate sensitivity experiments by increasing $CO_2$ at 1% per year until quadrupling
- ⊙ Abrupt $CO_2$ quadrupling experiment
- ⊙ Future projection for 21$^{st}$ century (2015-2100) as per the CMIP6 protocols,
- ⊙ Global Monsoon MIP (Model Intercomparison Project) for the period 1850-2014.

⊙ The CMIP6 data will be disseminated to all users using the Earth System Grid Federation (ESGF) node that has been setup at CCCR, IITM, Pune (Figure 3.10). The dissemination of CMIP6 data is targeted from 2019 through 2020.

⊙ The development of next generation IITM ESM, which would include interactive aerosols and chemistry and improved physical processes, is planned in the next 5 years. In addition, we plan to generate very high-resolution (grid size ~ 27 km) climate change simulations and future projections using the atmospheric-only component of the IITM ESMv2 in the next 3-4 years. The high-resolution climate change projections are important for assessment of changes in weather and climate extremes.

⊙ **High-resolution 20$^{th}$ century simulations and 21$^{st}$ century projections:** In addition to the IITM-ESMv2 Research and Development activities, the CCCR has generated high-resolution simulations of 20$^{th}$ century climatic variations and future projections using a global atmospheric model with telescopic zooming (~ 35 km in longitude x 35 km in latitude) over the South Asian region (Ref: Sabin et al. 2013, Krishnan et al. 2016). These high-resolution projections are useful to understand changes in monsoon rainfall, precipitation extremes, heat waves, droughts and floods, changes in cyclonic weather systems, hydrological cycle etc. Model outputs of the rainfall and temperature from the high-resolution simulations are made available for downloads http://cccr.tropmet.res.in/home/workshop/sept2014/index.html

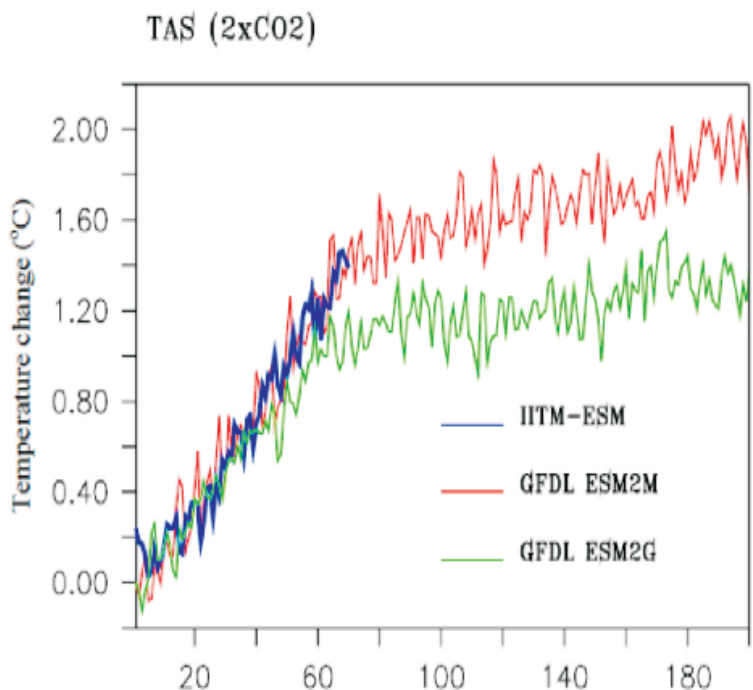

*Figure 3.8* Time-series of simulated global mean surface temperature (°C) anomalies based on the transient climate sensitivity experiments of the IITM-ESMv2 and the Geophysical Fluid Dynamics Laboratory (GFDL, USA) models (ESM2M and ESM2G) shown for comparison. The temperature anomalies are with respect to the PI control experiment. In the transient climate sensitivity experiment, CO2 is increased at a rate of 1% per year until doubling and thereafter it is kept constant. It can be noticed that the temperature increase in the IITM-ESMv2 is comparable with that of the GFDL-ESM2M, whereas the GFDL-ESM2G shows lower climate sensitivity. The IITM-ESMv2 and GFDL-ESM2M use the same ocean component, so that the oceanic heat uptake is similar in the two models, which may explain the similar response of global temperatures to increasing $CO_2$.

36

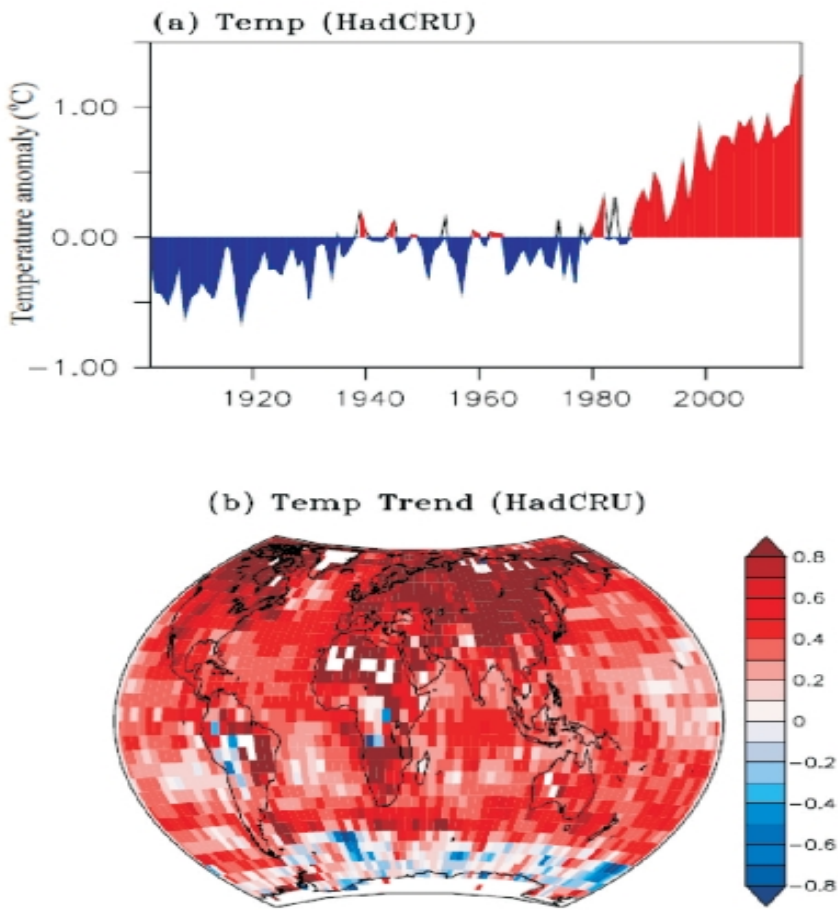

*Figure 3.9 (a)* Time-series of the observed global mean surface temperature (°C) anomalies for the period 1900-2016. A clear increasing trend [~ 0.85°C (116 years)$^{-1}$] in the global mean temperatures can be noted *(b)* Spatial map of linear trend in surface temperature over the period (1900-2016). Notice that the spatial pattern of warming is not uniform. The warming trend is stronger over the extra-tropical regions of North America, Europe, Asia, South America, Africa and Australia, and relatively smaller in magnitude over the tropical and near-equatorial areas. Special modeling experiments are included as part of the CMIP6 activity for detection, attribution and future projection of spatio-temporal variability of the climate change signal.

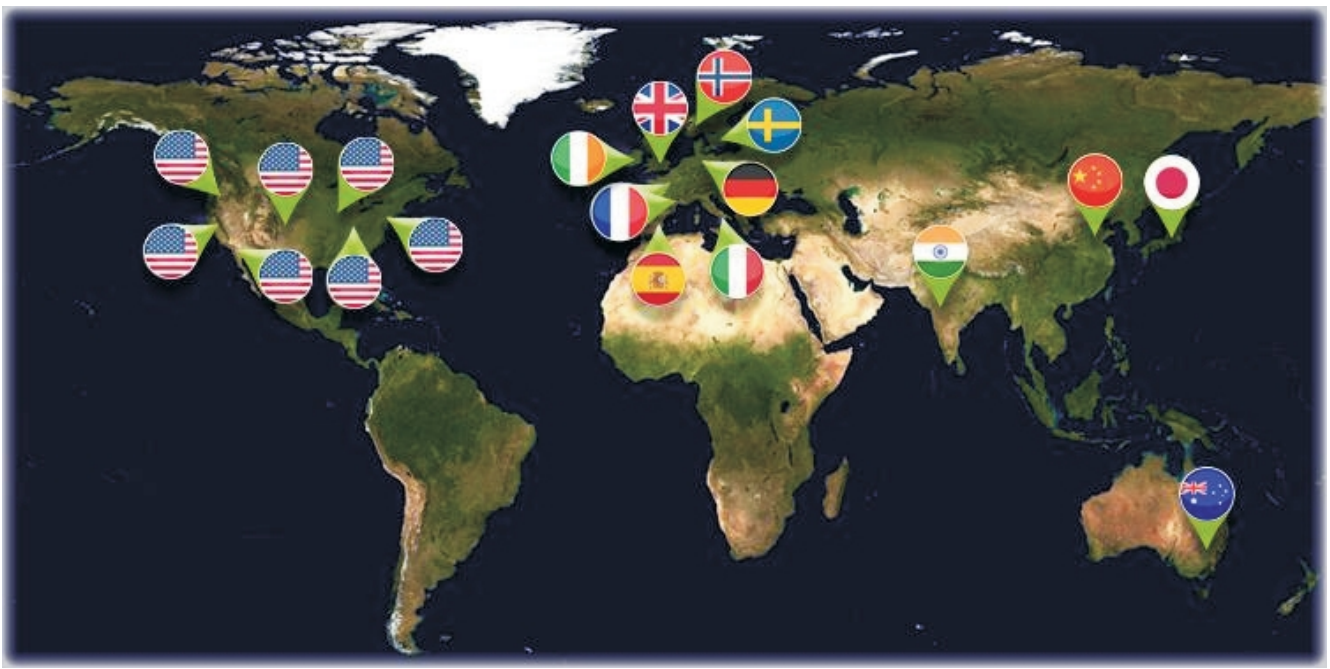

*Figure 3.10 Dissemination of CMIP6 and CORDEX South Asia datasets from CCCR, IITM, Pune:* The Earth System Grid Federation (ESGF) maintains a global system of federated data centers that allow access to the largest archive of climate data world-wide. The ESGF Data Node at CCCR-IITM is focused on supporting CCCR-IITM climate model datasets (eg., CORDEX-South Asia and CMIP6).